%% file: nmrlogic.tex
\pdfoutput=1
\documentclass{ijuc}

\usepackage{cite}
\usepackage[pdftex]{graphics}
\usepackage{amsmath, amssymb}
\usepackage{tikz}
\usetikzlibrary{calc}
\usepackage{siunitx}
\usepackage[version=4]{mhchem}
\usepackage[labelformat=empty]{subfig}

\newcommand{\negbar}[1]{\overline{#1}}
\newcommand{\XNOR}{\negbar{\oplus}}

\newcommand{\refsubonly}[1]{\protect\subref{#1}}

\let\oldSI\SI % For recursive naming without infinite loop (thx SO)
\renewcommand{\SI}[1]{\oldSI[separate-uncertainty = true]{#1}}

\newcommand{\vM}{$\mathbf{M}$ }

\newcommand{\absi}[1]{\ensuremath{\iota_{#1}}}

\begin{document}

\title{Discrete and Continuous Systems of Logic in Nuclear Magnetic Resonance}

\author{
Pedro M. Aguiar\inst{1}\email{pedro.aguiar@york.ac.uk} \and
Robert Hornby\inst{2}\email{robjhornby@gmail.com} \and
Cameron McGarry\inst{2}\email{cameron@cmcgarry.co.uk} \and
Simon O'Keefe \inst{3}\inst{4}\email{simon.okeefe@york.ac.uk} \and
Angelika Sebald\inst{1}\inst{4}\email{angelika.sebald@york.ac.uk}
}

\institute{
Department of Chemistry, University of York, York YO10 5DD, UK \and
Corpus Christi College, Merton Street, Oxford OX1 4JF, UK \and
Department of Computer Science, University of York, York YO10 5DD, UK \and
York Centre for Complex Systems Analysis, University of York, York YO10 5DD, UK
}

\def\received{Received \today; In final form \today} %TODO

\maketitle

\begin{abstract}
We implement several non-binary logic systems using the spin dynamics of nuclear
spins in nuclear magnetic resonance (NMR). The NMR system is a suitable test
system because of its high degree of experimental control; findings from NMR
implementations are relevant for other computational platforms exploiting
particles with spin, such as electrons or photons. While we do not expect the
NMR system to become a practical computational device, it is uniquely useful to
explore strengths and weaknesses of unconventional computational approaches,
such as non-binary logic.
\end{abstract}


\section{Introduction}

Implementations of computations on less conventional platforms such as DNA
\cite{Amos2002}, slime moulds \cite{Whiting2014}, oscillating chemical reactions
\cite{Gorecki2015} or liquid crystal media \cite{Harding2005} have recently seen
increased activities and attention with a view to exploring new ideas in the
theory of logic gates. However, despite the unconventional nature of the
computational platform, the form of computation in the vast majority of cases is
based on a binary representation and binary logic.

The predominance of binary logic in computation is at least partly a consequence
of previous choices of technology to implement computation, which worked well
for binary. There are many examples of non-binary computational machines,
including Babbage's Difference and Analytic engines which used denary
\cite{Swade2002}, and a wide variety of analogue systems \cite{Murrell2013}.
Consideration of new, less conventional, technologies allows us to
reconsider what we implement and look again at other computational systems, such
as those with an unconventional base. The dynamics of nuclear spins have been
explored in this context, building on previous work that looked at nuclear
magnetic resonance (NMR) systems and binary logic \cite{Bechmann2012}.

In a computational context, the NMR system and its spin dynamics are probably
more widely known for their applications in quantum computing \cite{Jones 2011}.
However, nuclear spin dynamics have also played a role as an extremely
versatile and highly controllable experimental platform in the implementation of
classical computations \cite{Bechmann2012}, highlighting the advantages of the
NMR system as a sandpit for design-oriented theoretical work or as a
developmental tool for, for example, optical computation.

Previously we have taken an NMR-based design approach for the implementation of
binary logic gates \cite{Bechmann2012}. Combined consideration of theoretical
descriptions of binary logic gates as well as the NMR properties of (simple)
nuclear spin systems in the liquid state lead to a number of suggestions. For
example, an NMR-focussed starting point suggests that a ternary logic system
\cite{Glusker2005, Brusentsov2011} might make better use of the natural
occurrence of values \{-1, 0, +1\} in a system made up of an ensemble of
spin-$1/2$ particles than does binary logic.  From the starting
point of mathematical logic, it appears attractive to investigate properties of
logic based on complex numbers, and how this maps to possible experimental NMR
implementations.

\section{Summary of NMR experiments}

\input{./sections/nmrsummary}

\section{Ternary Logic}
\label{ternary}

\input{./sections/ternary}

\section{Continuous Complex-Number-Based Logic}
\label{complex}

\input{./sections/complex}

\section{Experimental}

\input{./sections/experimental}

\section{Outlook}

\input{./sections/outlook}

\section{Acknowledgements}

We gratefully acknowledge support of our work by the York Centre for Complex
Systems Analysis (YCCSA) summer schools 2014 and 2015.

\appendix

\section{Ternary Classes}

\input{./sections/ternaryclasses}

\section{Generalised Continuous Logic}

\input{./sections/generalisation}

\bibliography{nmrlogic}{}

\end{document}

%% file: sections/nmrsummary.tex
Before discussing ternary and complex-number based logic in the context of NMR
implementations we briefly discuss some of the basic underlying principles of
NMR experiments underpinning our work; we will use nomenclature
introduced in earlier work \cite{Bechmann2012}. Our work is restricted to some
of the most simple NMR experiments.

\begin{figure}[h!t]
\centering
\includegraphics[width=\linewidth]{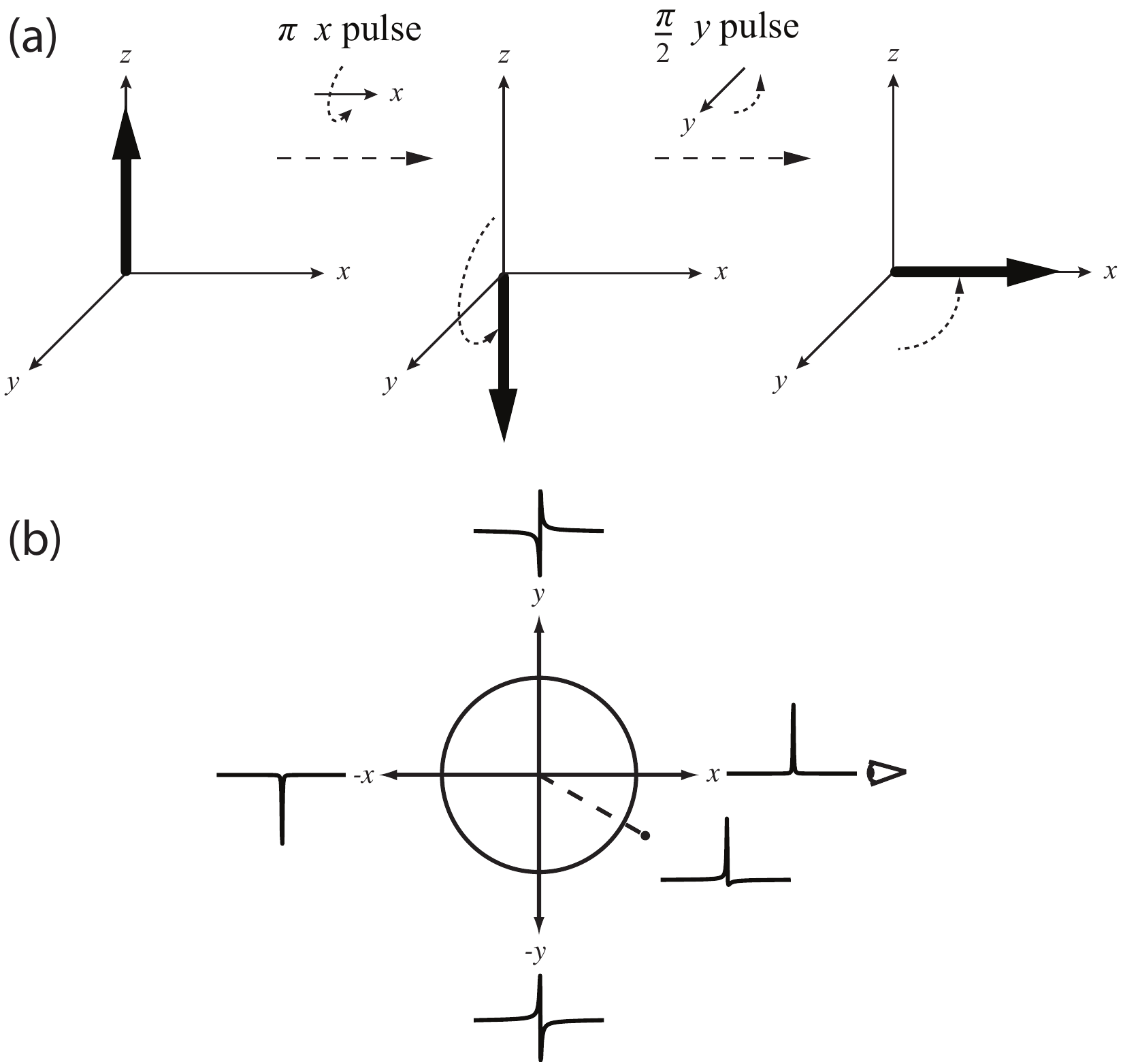}
\caption{
(a) Illustration of r.f.pulses rotating the bulk magnetisation vector away
from its equilibrium $+z$ orientation.
(b) The magnitudes and phases of NMR signals after Fourier transformation of the
time domain signals. Note that both r.f. pulses and observation can be on or off
resonance \cite{Levitt2008}}.
\label{fig:sum}
\end{figure}

Our samples are simple liquid compounds representing $^1\mathrm{H}$ nuclear spin
systems that can be fully described by their bulk magnetisation vectors. The
effects of radiofrequency (r.f.) pulses on the magnetisation vectors can be most
easily visualised as (positive) rotations of the vector by a specified angle
around a particular axis. This is illustrated in Figure \ref{fig:sum}(a).

Also shown are the corresponding NMR signals in the frequency domain (Fourier
transformation of the observed time domain signal) carrying magnitude and phase
information. Recall that observation of the bulk NMR signal is always the
projection of the magnetisation vector onto the $x$-$y$ plane.  Figure
\ref{fig:sum}(b) highlights the choice of signal amplitudes and phases as the
basis for construction of ternary logic gates. Note that both off- and
on-resonance options exist \cite{Levitt2008} for excitation as well as
observation. Figure \ref{fig:sum}(a) highlights the choice of signal amplitudes
\emph{and} phases as the basic ingredients for construction of logic gates.

%% file: sections/ternary.tex
A ternary logic function with two input variables mapping to one output value
can be described by a $3\times3$ truth table, shown in Table
\ref{table:ternarytable}. Using the logic values of the balanced ternary system,
\{-1, 0, 1\}, is a straightforward choice for the range of NMR experiments being
considered, in which these three values occur naturally.

\begin{table}

\begin{center}

\begin{tabular}{ r r | r r r }
& & & \textbf{B} & \\
& & -1 & 0 & 1 \\
\hline
&-1 	& 1 	& 0 	& -1 \\
\textbf{A}&0 	& 0 	& 0 	& 0 \\
&1 	& -1 	& 0	& 1 \\
\end{tabular}

\caption{An example of a ternary logic function - ternary multiplication, with
input A on the left, input B above, and the nine output values corresponding to
the nine pairs of input values in the main part of the table.}
\label{table:ternarytable}
\end{center}
\end{table}

Each of the nine pairs of input values can lead to any of three output values,
meaning there are $3^9 = 19{,}683$ possible functions of this sort. When
attempting to implement a ternary logic function in a physical system or,
inversely, trying to work out which logic functions may be implemented by a
given physical system, it may be useful to classify these functions based on
physical equivalences to reduce the search space; if two or more functions can
be represented by an identical physical system, they need not be considered
separately. These equivalences come about because any given physical
implementation of a logic gate could be reinterpreted as another logic gate just
by remapping the physical values of the implementation to different logic
values.

For example, a binary logic gate can be represented by an electronic circuit,
where a high voltage is generally chosen to correspond to a 1, but could just as
easily be chosen to correspond to a 0 (with a low voltage corresponding to the
other value in each case). By making use of this freedom of relabelling, more
than one logic gate could be represented by the same electronic circuit.

The approach of canalising inputs taken in earlier work on binary logic
\cite{Bechmann2012} puts all two-input, one-output binary logic gates into four
classes based on patterns in the parameters of the experiments implementing
those logic gates. These four classes are the same as those produced by the
Negation-Permutation-Negation (NPN) classification system\cite{Correia2001}.

In the NPN classification, two binary logic functions are considered to be in
the same class if one can be converted into the other by negating input values,
permuting the order of the input values, or negating the output values, in any
combination. This amounts to the same thing as remapping physical values to all
the possible different combinations of logic values, and so each one of these
classes is a list of logic functions which can be implemented by the same
physical system.

These two approaches to classifying logic gates have different ease of
application in different circumstances, one being parameter-centric which is
useful when looking at a physical system, and the other being based on
mathematical transformations, but both lead to the classes which will aid the
search for logic gate implementations. We extend the ideas of the
parameter-centric canalising approach to ternary logic so that searching for
implementations of ternary logic functions may be made easier.

The canalising input values approach determines which NPN classes can be
implemented by a physical system by looking at how the parameters of that system
behave. Specifically, it looks for values of parameters which canalise the
output of the system, i.e. in a two input, one output binary logic gate, if
there is a certain value for one of the inputs for which the output is a
constant value, then that input value is said to be canalising. By determining the
behaviour of the parameters in this way, the class of logic functions which
could be mapped on to that system are found immediately.

An extension of the NPN classification of binary logic functions places the
19,683 different two-input, one-output ternary logic functions into 84 different
equivalence classes\cite{Lloris1981, Soma2011} by still allowing the order
of the inputs to be swapped, and replacing the single negation function of
binary logic with the six permutation functions of ternary logic. See Appendix
\ref{ternaryclasses} for details of this classification system.

\begin{figure}
\centering
\includegraphics[width=\linewidth]{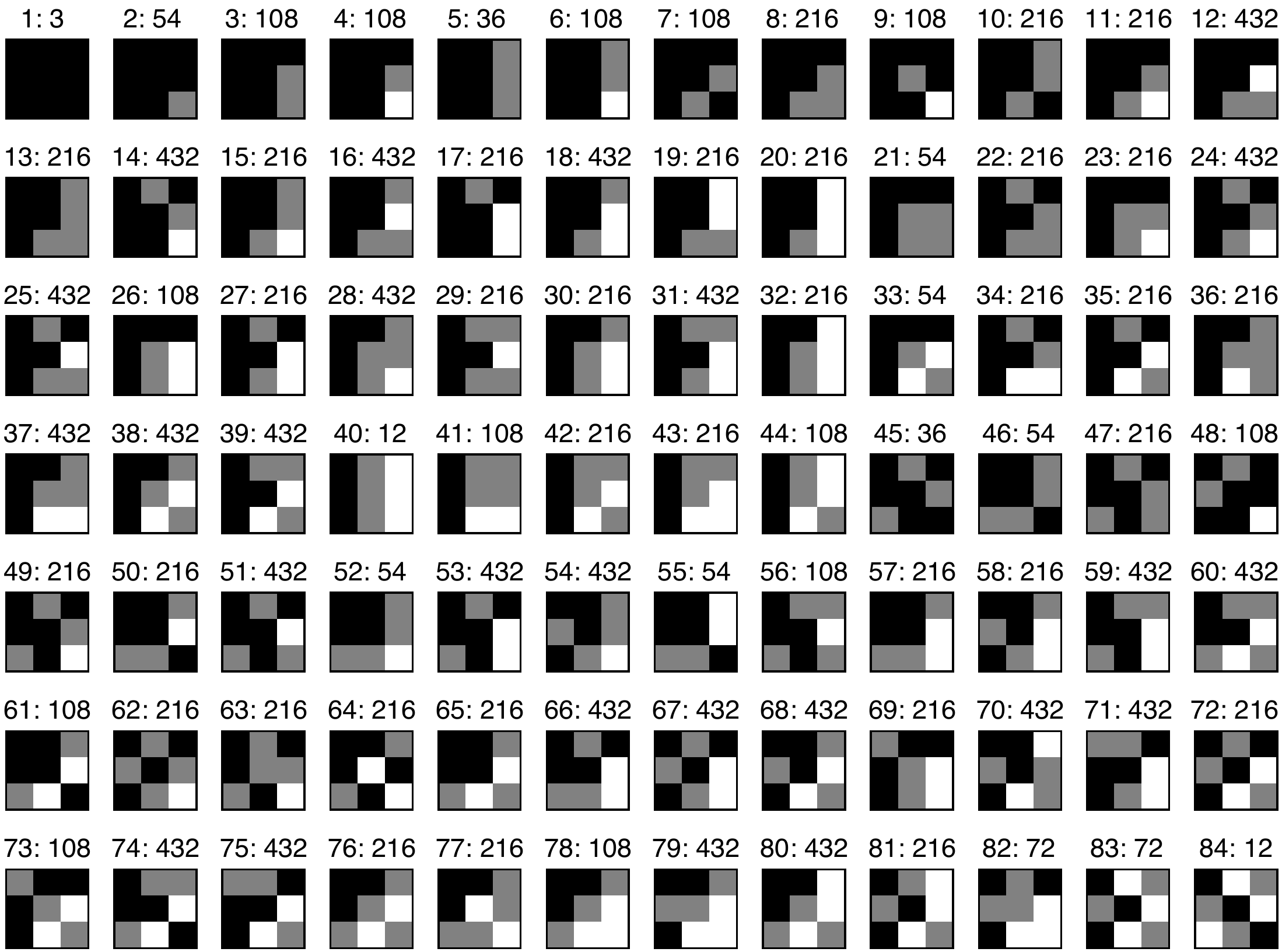}
\caption{
The 84 canonical logic functions with two inputs and one output which represent
each of the 84 equivalence classes in this type of logic function. These
functions can be transformed under the equivalences described in Appendix
\ref{ternaryclasses} to produce all 19,683 functions of this type. Each class is
numbered based on the order found in previous classifications\cite{Lloris1981,
Soma2011}. The number of functions present in each class is given following the
class number.}
\label{fig:allclass}
\end{figure}

The algorithms previously used for classifying a ternary logic function do not
greatly simplify the search for implementations of logic gates in novel
substrates because they do not obviously relate to the behaviour of parameters
in physical systems. One algorithm \cite{Lloris1981}, for example, takes a given
logic function and transforms it into a canonical logic function which
represents the class it belongs to, which does not translate to features of a
physical system in a straightforward way. All of these canonical logic functions
are shown in Figure \ref{fig:allclass}.

By looking at the ternary NPN classes, it is apparent that certain features
which resemble the canalising inputs of binary logic functions are present
within any given class, though a simple parameter-centric classification from
these features has yet to be found which matches the NPN classification exactly.

A classification which contains a mixture of individual NPN classes and unions
of two or more NPN classes is found when one considers a set of measures which
reproduce the canalising inputs classification when applied to a binary logic
gate, but can also be applied to a ternary logic gate:

\begin{itemize}
\item The number of different output values in each column, with
order unimportant between columns,
\item The number of different output values
in each row, with order unimportant between rows.
\end{itemize}

These two measures can be taken in either order, and then every gate which
has a matching set of measures is in the same parameter-centric class (PC class).

\begin{figure} \begin{center}
\scalebox{0.4}{\includegraphics{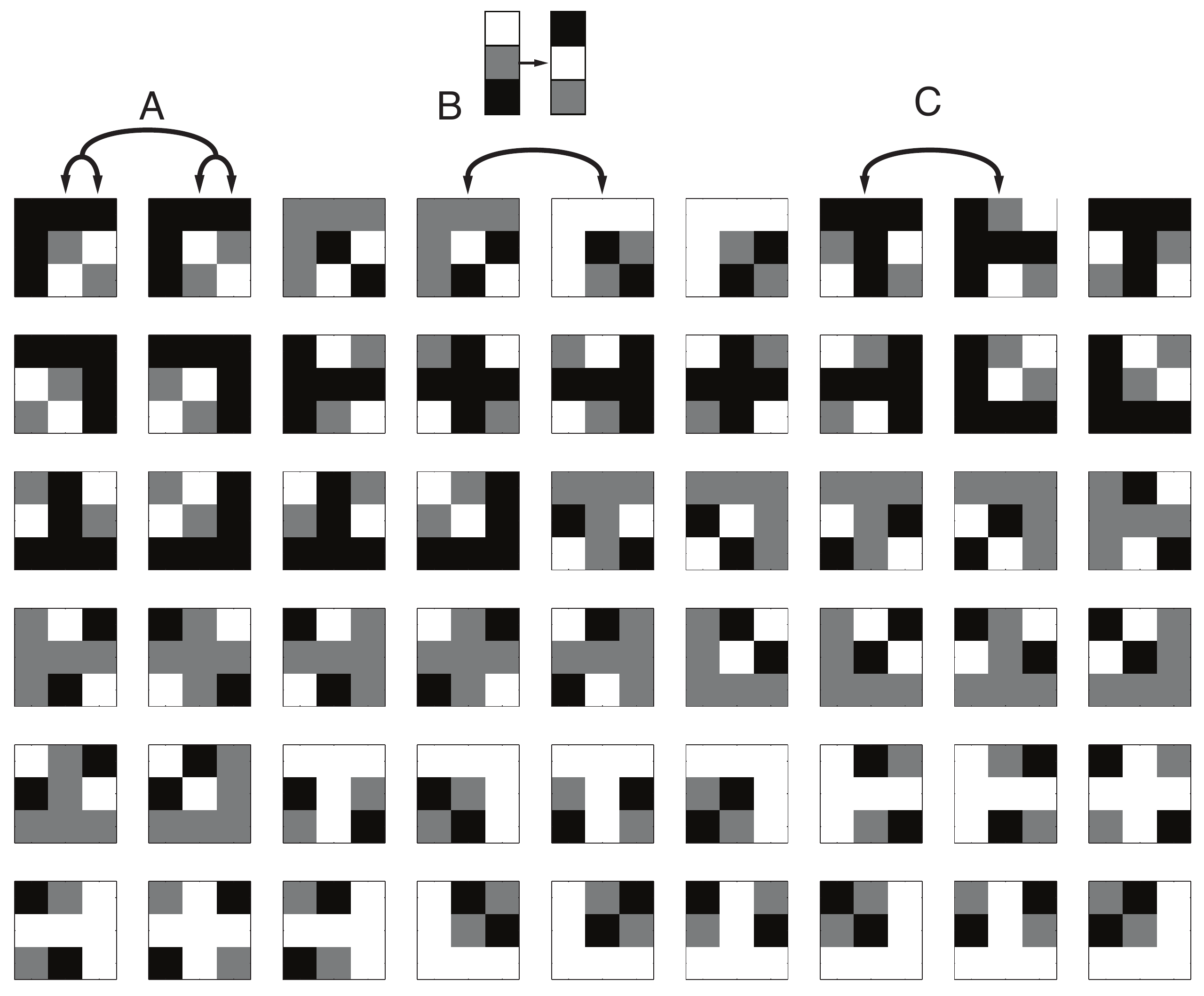}} \end{center}
\caption{
The 54 ternary logic functions which make up the NPN equivalence class which
contains ternary multiplication. The three equivalences are exemplified.  A:
Swapping the order of the second two columns (relabelling $\{0,
1\}\rightarrow\{1,0\}$ in the top input), B: permuting the output values
(relabelling $\{-1,0,1\}\rightarrow\{1,-1,0\}$ as shown above), C: swapping the
order of the inputs - a reflection about the main diagonal. A physical system
which implements one of these logic functions could be
relabelled to represent any of these functions.}
\label{fig:class33}
\end{figure}

As an example, the PC class which contains the ternary
multiplication function, shown in Figure \ref{fig:class33}, will have one row
(or column) with a constant output value, and the other two rows (or columns)
will have three different output values. In addition, one column (or row) will
have a constant output value, and the other two columns (or rows) will have
three different output values.

Some NPN classes share the same set of measures in the PC classification, and so
some PC classes are a union of two or more NPN classes. The PC classification
makes finding an implementation for a ternary logic function contained in one of
the PC classes which are equal to a single NPN class more straightforward, and
still narrows down the search when the ternary logic function to be implemented
is in one of the PC classes with NPN overlap, although the overlap does add some
of the complication back in. Adding further measures to the PC classification
could separate the PC classes to be the same as the NPN classes at the cost of
the simplicity of the measures.

\subsection{NMR Implementations}

\begin{figure}[ht]
\centering
\includegraphics{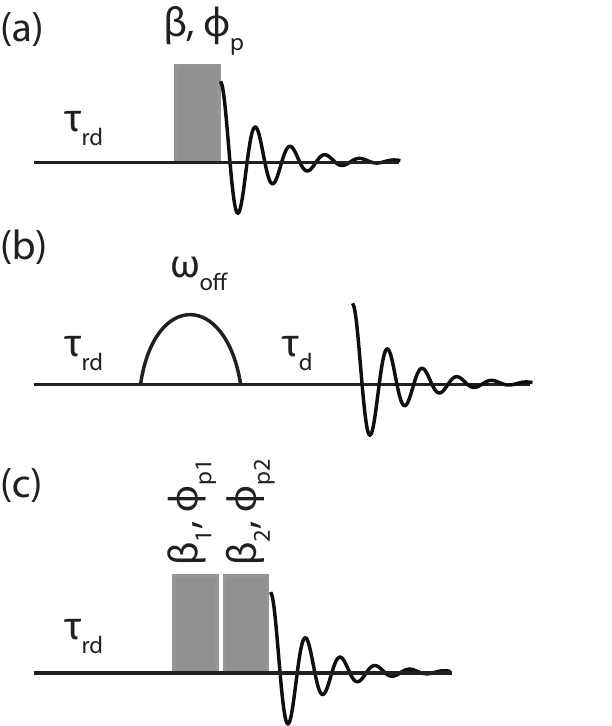}
\caption{
Pulse sequences used for ternary logic experiments. (a) A single square pulse of
variable duration (rotation) and phase, (b) A frequency-selective pulse is
followed by a delay before acquisition and (c) two square pulses each of
adjustable duration (rotation) and phase. All experiments are preceded by a
suitably long delay ($\tau_\mathrm{rd}$) to ensure the system is at equilibrium.
The arched pulse in (b) is selective.}
\label{fig:ternsequences}
\end{figure}

One possible NMR implementation of a ternary logic gate can be found in an
experiment with a single frequency in the spectrum, using a single pulse with
flip angle, $\beta$, and the phase, $\phi_p$, as the two input parameters. This
pulse sequence is shown in Figure \ref{fig:ternsequences}(a).

\begin{figure}[h!t]
\centering
\includegraphics[width=\linewidth]{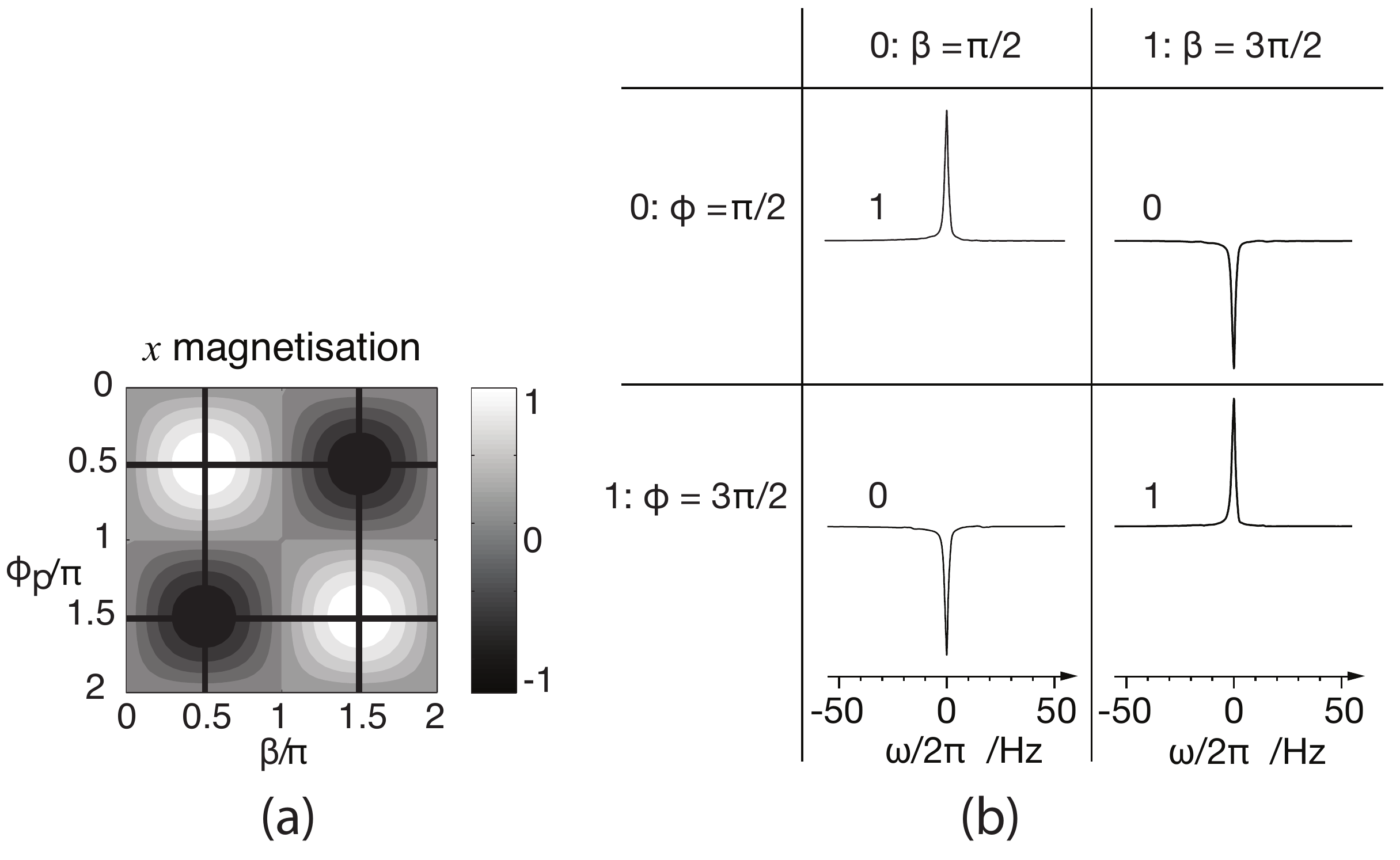}
\caption{
(a) A contour plot showing the expected $x$ magnetisation for a single
pulse experiment as a function of the phase of the pulse, $\phi_p$, and flip
angle, $\beta$. (b) By performing this experiment with the $\phi_p$ and $\beta$
values shown, this experiment can be interpreted as the binary XOR logic
function, with the integral of the frequency domain signal corresponding to the
$x$ magnetisation.}
\label{fig:bphi_bin}
\end{figure}

A contour plot of the expected resultant magnetisation is shown in Figure
\ref{fig:bphi_bin}(a). Previously, this setup has been
used to implement a binary XOR gate - the chosen parameter values which
represent an XOR gate are shown in the figure, and the result of performing the
experiment with each combination of input values is also shown.

In this plot, lines have been drawn through the chosen parameter values, so that
the PC measures can be found. There is a horizontal section of constant value
and a vertical section of constant value, as well as two other horizontal and
vertical sections which go through positive, negative and zero values. These
measures define a PC class which corresponds to just one NPN class - the class
which contains ternary multiplication.

\begin{figure}[h!t]
\centering
\includegraphics[width=\linewidth]{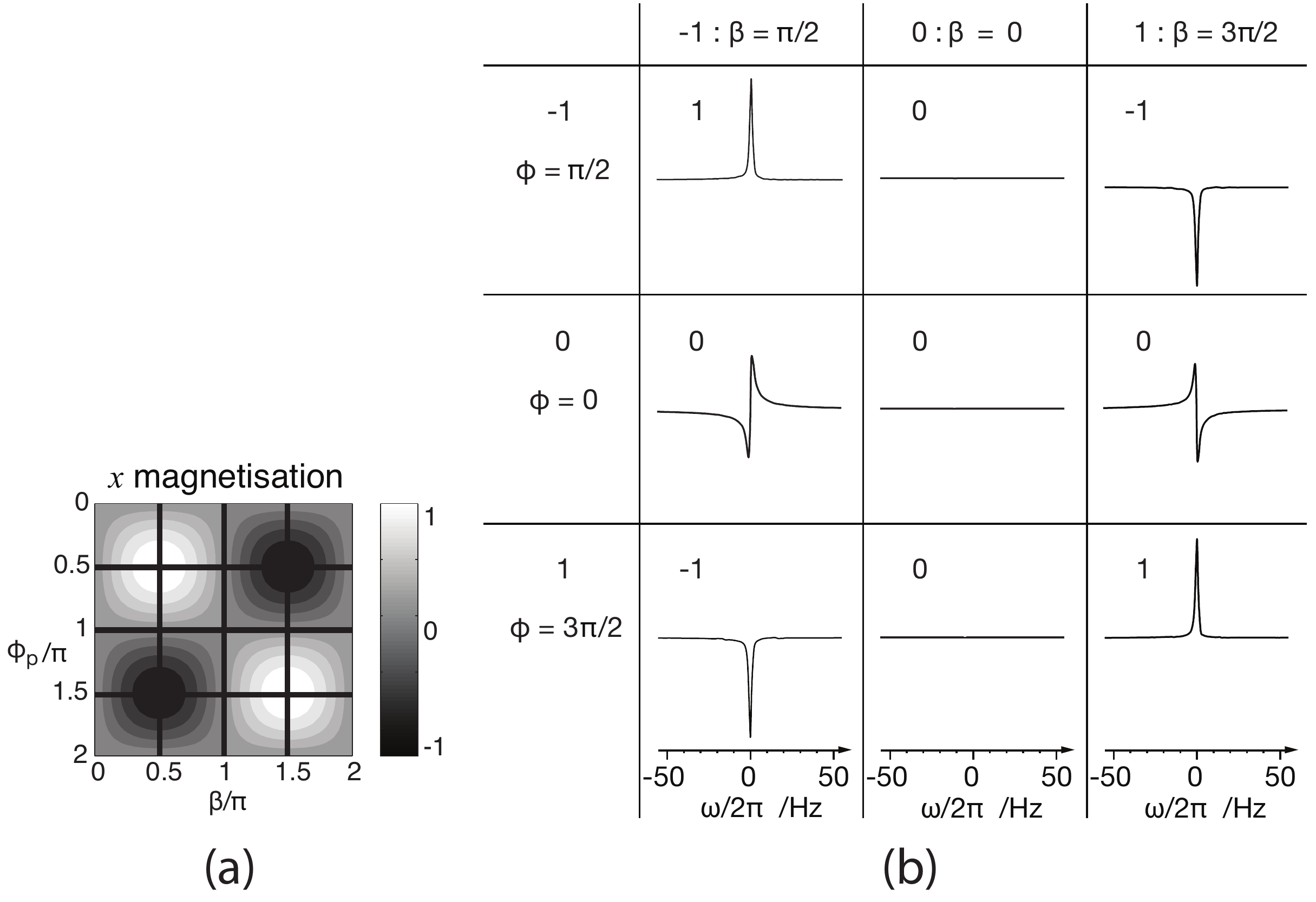}
\caption{
An implementation of the ternary multiplication function is shown, taking
different $\phi_p$ and $\beta$ values from the same experiment as in Figure
\ref{fig:bphi_bin}, shown in (a). The result of each possible combination of
input values is laid out as a ternary logic table in (b).
}
\label{fig:bphi_tern}
\end{figure}

If more than one NPN class corresponded to this parameter-centric
classification, we would need to make further checks to confirm that ternary
multiplication can be implemented, but because only one class corresponds to it,
we can immediately conclude that multiplication can be implemented by this
experiment, as is shown in Figure
\ref{fig:bphi_tern}(b).

\begin{figure}[h!t]
\centering
\includegraphics[width=\linewidth]{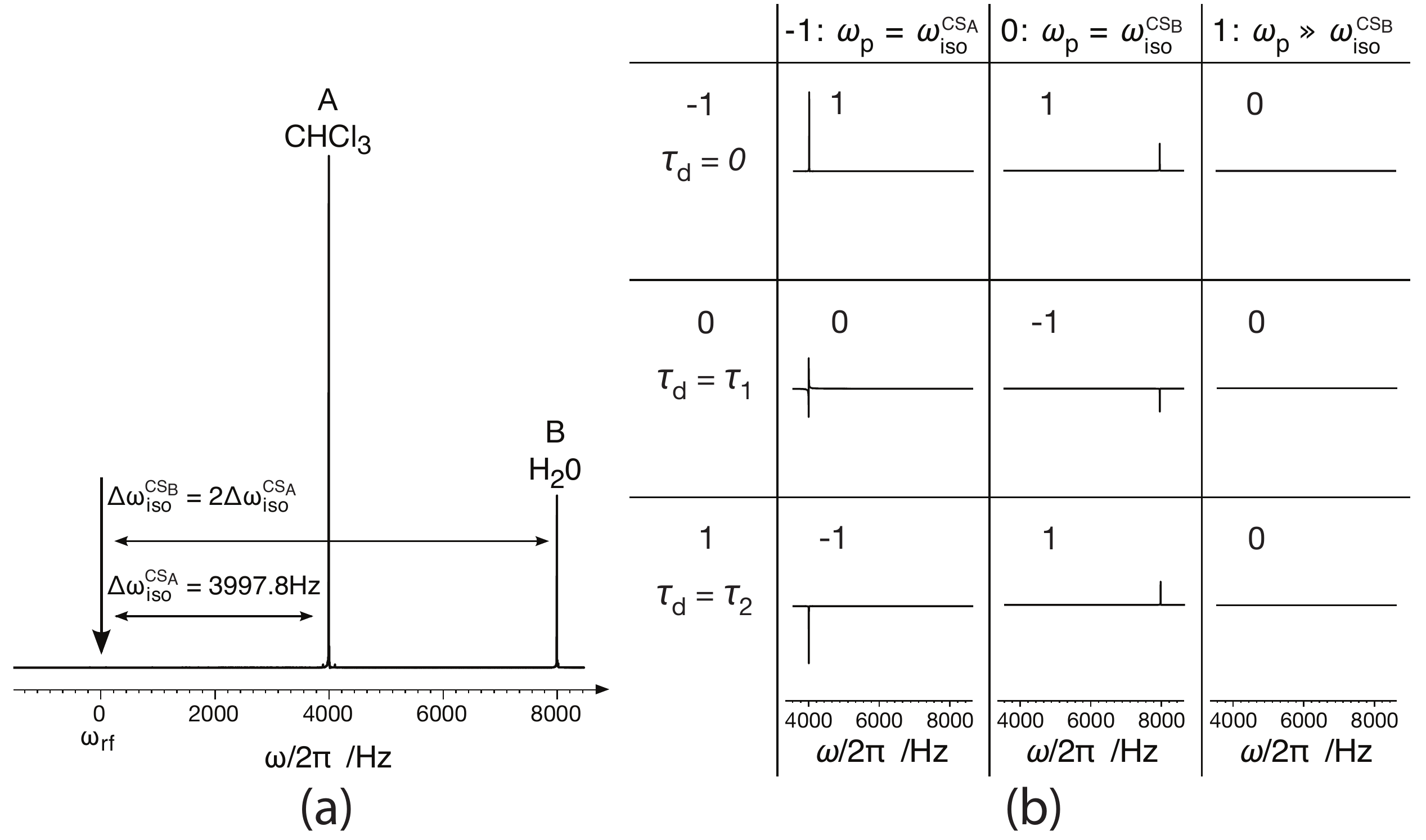}

\caption{
(a) The sample spectrum used for the selective delay implementation, along
with the transmitter frequency $\omega_{\text{rf}}$, selected so that peak B has
twice the frequency difference from the transmitter as peak A, so that the spins
of B will have precessed twice as far as the spins of A in a time delay before
acquisition, $\tau_d$. (b) The result of using the pulse sequence in
Figure \ref{fig:ternsequences}(b) with the selected values for $\tau_d$ and
$\omega_p$ shown. $\tau_1$ corresponds to an acquisition delay which allows the
spins of B to precess by $\pi$ (the spins of A precess by $\pi/2$), and $\tau_1$
allows the spins of A to precess by $\pi$ (the spins of B precess by $2\pi$).
The selective frequency of the pulse, $\omega_p$, leads to a signal from either
A, B, or neither. The integral of the resulting frequency domain signal
corresponds to the logic values shown.  }
\label{fig:selectdelay}
\end{figure}

Another NMR implementation could make use of a sample with peaks at more than
one frequency, as shown in Figure \ref{fig:selectdelay}(a).  This allows for a
pulse sequence involving selective pulses and delays, see Figure
\ref{fig:ternsequences}(b). This allows two more possible parameters of the NMR
experiment to be used to find logic gates with different parameter behaviours to
those found in the single spectral peak experiment. The abundance of parameters
and freedom to keep expanding the experiment strongly suggests that any of the
84 NPN classes of logic gates should be implementable by NMR.

One such implementation using two spectral peaks was tested, varying the
frequency of the selective pulse as one input parameter and the time delay
before acquisition as the second parameter. The results of this experiment are
shown in Figure \ref{fig:selectdelay}(b), laid out as a ternary logic function
table. The PC class of this experiment (and therefore the corresponding NPN
classes) is one which couldn't be achieved by the single pulse, single spectral
frequency experiment.

To begin exploring beyond ternary logic, an experiment involving two pulses was
performed on a sample with a single spectral peak. The pulse sequence for this
experiment is shown in Figure \ref{fig:ternsequences}(c).

Two of the pulse parameters were varied sequentially: the flip angle, $\beta_1$,
of pulse one, and the phase, $\phi_{p2}$ of pulse two. The other parameters were
set to fixed values, $\phi_{p1} = \frac{\pi}{2}$ and $\beta_2 =
\frac{\pi}{2}$. 100 results were taken, in a 10x10 grid with values of
$\beta_1$ taken evenly spaced over the interval $[0, 2\pi]$ and $\phi_{p2}$ also
evenly sampled over $[0, 2\pi]$.

\begin{figure}
\centering
\includegraphics[width=\linewidth]{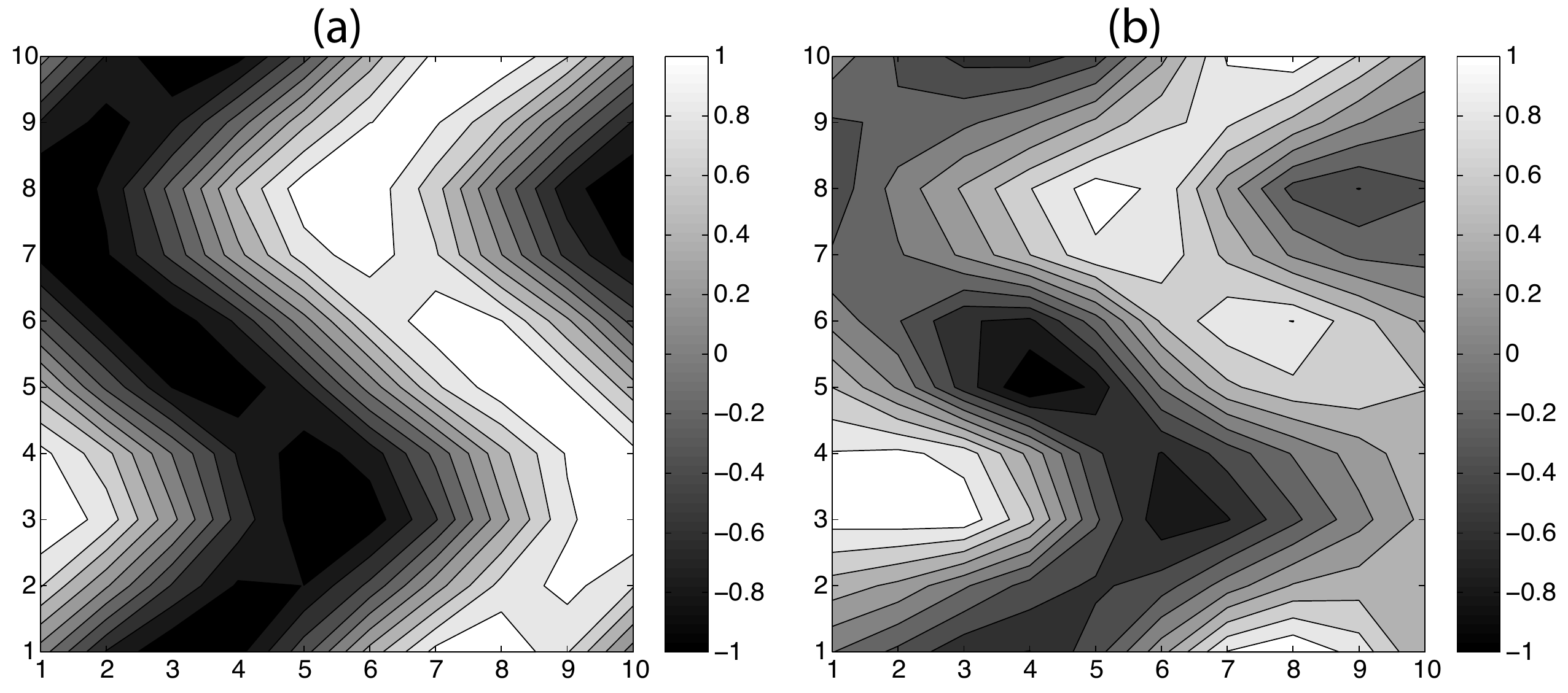}
\caption{
The comparison of the expected (a) and measured (b) results of
a 10$\times$10 array of NMR experiments using the pulse sequence in Figure
\ref{fig:ternsequences}(c). The the flip angle of pulse one, $\beta_1$, was varied
over the range $[0,2\pi]$ in the $x$ direction and the phase of pulse two,
$\phi_{p2}$ was varied over the range $[0,2\pi]$ in the $y$ direction, with
fixed $\phi_{p1} = \frac{3\pi}{2}$ and $\beta_2 = \frac{\pi}{2}$.}
\label{fig:2pulsesequence}
\end{figure}

These data were compared to a theoretical model and were shown to match the
expected results to within a small error, showing that the system of predicting
NMR experimental outcomes is robust. The comparison of the theoretical results
and the measured results is shown in Figure \ref{fig:2pulsesequence}. This
array of values leads to the idea of continuous logic.

%% file: sections/complex.tex
Rather than using the discrete truth-values found in Boolean algebra, it has
been established that a system of continuous logic can be implemented that uses
continuous truth-values to express uncertainty \cite{Levin2000}. In an attempt
to extend upon the ideas of the previous section we construct Complex Logic, a
system of logic that has been split into two parts based on the exponential form
of complex numbers: Magnitude Logic (mLogic) and Phase Logic (pLogic).

When representing information as a complex number, we can arbitraily define any
function on $n$ complex numbers in terms of two other functions

\begin{equation}
\label{eqn:generalop}
f(z_1, \cdots z_n) = g(r_1, \cdots r_n)\exp(ih(\theta_1, \cdots \theta_n))
\end{equation}

where $z_k = r_ke^{i\theta_k}$. The following subsections focus on the ways in
which these functions can be defined in order to produce a meaningful logic. The
possible choices for $g$ are described by mLogic and those for $h$ by pLogic.
A more mathematically rigorous approach to continuous logic can be found in
appendix \ref{app:general}.

\subsection{Magnitude Logic}

Fuzzy logic is a form of continuous logic that has been well-developed elsewhere
\cite{Levin2000}. Here we state that all aspects of mLogic corresponds in
some way to fuzzy logic. For this to be the case it is necessary that only
complex numbers with magnitude $r \in [0,1]$ are considered.

When refering to the comparison between mLogic and pLogic with fuzzy logic, the
operations in Fuzzy Logic will be refered to as ``normal fuzzy logic
operators'' and denoted mathematically by a subscript $0$. mLogic and pLogic
operators are denoted by a subscript $m$ and $p$ respectively.

For any mLogic operator acting on a complex number, the result will have the
same phase but with a magnitude defined by the  normal fuzzy logic operator. For
some operator $F_m$, the fuzzy logic match is $F_0$ such that

\begin{equation}
F_m(re^{i\theta}) = F_0(r) e^{i\theta}.
\label{eqn:mfnctun}
\end{equation}

One example is the unary operator NOT where $\neg_0 x = 1 - x$, it
follows then that

\begin{equation}
\neg_m(re^{i\theta}) = (1 - r)e^{i\theta}.
\label{eqn:mNOT}
\end{equation}

More generally, for an operation that is not necessarily unary

\begin{equation}
F_m(z_, \cdots z_n) = F_0(r_1, \cdots r_n)e^{i\theta_\mathrm{res}}
\label{eqn:mfunct}
\end{equation}

where $z_k = r_ke^{i\theta_k}$ and $\theta_\mathrm{res} =
\theta_\mathrm{res}(\theta_1, \cdots \theta_n)$ can be defined arbitrarily.
The ways in which $\theta_\mathrm{res}$ can be defined are discussed in
section \ref{sec:total}.

For example AND ($\wedge$) with $x \wedge_0 y = xy$ gives

\begin{equation}
z_1 \wedge_m z_2 = r_1r_2e^{i\theta_\mathrm{res}}
\label{eqn:mAND}
\end{equation}

where $\theta_\mathrm{res} = \theta_\mathrm{res}(\theta_1, \theta_2)$ must be
defined separately as in equation \ref{eqn:generalop}.

\begin{figure}[h!tb]
\centering
\includegraphics{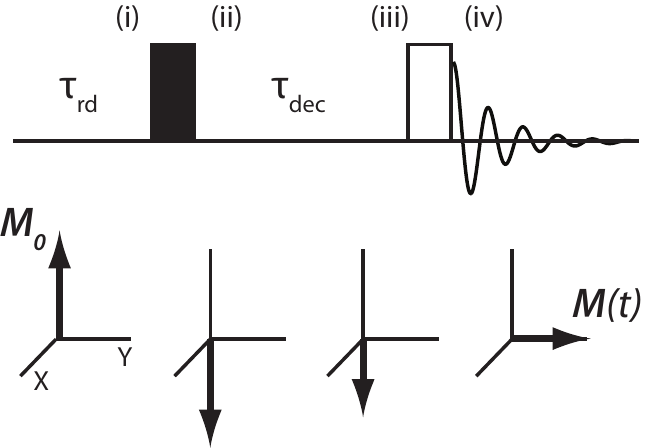}
\caption{
A pulse sequence capable of encoding magnitude along with a schematic of the
evolution of the net magnetisation \vM throughout the sequence at the points (i)
through (iv). Filled pulses represent rotations of the magnetisation by $\pi$
and unfilled by $\frac{\pi}{2}$.}
\label{fig:magpulse}
\end{figure}

The representation of a magnitude in the NMR system is achieved by taking
advantage of the $T_1$ decay \cite{Levitt2008} as illustrated in Figure
\ref{fig:magpulse}.  We begin from equilibrium magnetisation (that is, \vM
aligned along the $+z$ direction, parallel to the external magnetic field) and
apply a $\pi$ pulse after which \vM will be aligned along $-z$. \vM immediately
begins to decay back towards the equilibrium position over time $t$ according
to

\begin{equation}
\label{eqn:sum:decay}
\mathbf{M}(t) = M(0)\left[1-2e^{-t/T_1}\right]\mathbf{\hat{z}}.
\end{equation}

We are free to choose the delay $t = \tau_\mathrm{dec}$ such that the magnitude
$r$ is described by the ratio $\alpha M(t)/M(0)$. The constant $\alpha$ can be
chosen arbitraily in order to present a range of magnitudes $r \in [0,\alpha]$.
In the case of mLogic $\alpha = 1$. Depending on how much time one allows for
this relaxation process, different magnitudes of the magnetisation vector can be
read out at different times.

A $(-\frac{\pi}{2})_y$ pulse is then applied so that \vM lies along the positive
$x$-axis.  The intensity of the output signal is used to calulate the magnitude.
The pulse sequence and corresponding movement of the vector are described in
Figure \ref{fig:magpulse}.  Such an experiment is more commonly used to
determine the value of $T_1$ \cite{Levitt2008}.

\subsection{Phase Logic}

When dealing with the phase of a complex number, we must take into account that
generally $e^{i\theta} = e^{i(\theta + 2\pi n)}$. For this reason it is defined
that all phases shall be expressed with $\theta \in [0,2\pi)$ and all arithmetic
of phases is modulo $2\pi$.

In pLogic, it is defined that $0$ represents truth and $\pi$ falsehood. There
are then two domains in which no member has absolute truth or absolute
falsehood: $\Theta_1 = (0, \pi)$ and $\Theta_2 = (\pi, 2\pi)$. There is a
relation between the truth values in pLogic and those in fuzzy logic: for any
truth value defined in pLogic (a pTruth value) $\theta$, there is an
equivalent truth value in fuzzy logic given by the projection
function.

\begin{equation}
T_0(\theta) = \frac{\lvert \pi - \theta \rvert}{\pi}.
\label{eqn:proj}
\end{equation}

\begin{figure}[h!tb]
\centering
\begin{tikzpicture}[scale=2.0]
  % Axes
  \draw [->] (-1.4,0) -- (1.4,0) node [above right] {$\Re\{z\}$};
  \draw [->] (0,-1.4) -- (0,1.4) node [above left] {$\Im\{z\}$};
  \draw [->] (-1.4,-1.65) -- (1.4,-1.65) node [above] {pTruth (non-linear)};

  % Axis Labels
  \filldraw[black] (1,0) circle(1pt) node [above right] {$1$};
  \filldraw[black] (-1,0) circle(1pt) node [above left] {$-1$};
  % \filldraw[black] (0,1) circle(0.4pt) node [above right] {$i$};
  % \filldraw[black] (0,-1) circle(0.4pt) node [below right] {$-i$};

  % Circle
  \draw[thick] (1,0) arc (0:180:1);
  \draw[very thick, dashed] (-1,0) arc (180:360:1);

  % Points
  % Circle
  \coordinate (O) at (0,0);
  \coordinate (z) at (65:1);
  \coordinate (z*) at (295:1);
  \coordinate (-z*) at (115:1);
  \coordinate(-z) at (245:1);
  %Truth axis
  \coordinate (pFalse) at (-1,-1.65);
  \coordinate (pTrue) at (1,-1.65);
  % Projections onto truth axis
  \coordinate (phi) at ($(pFalse)!(z)!(pTrue)$);
  \coordinate (-phi) at ($(pFalse)!(-z*)!(pTrue)$);
  % Draw nodes
  \filldraw[black] (z) circle(0.5pt) node [above right] {$z=e^{i\phi}$};
  \filldraw[black] (z*) circle(0.5pt) node [below right] {$z^*$};
  \filldraw[black] (-z*) circle(0.5pt) node [above left] {$-z^*$};
  \filldraw[black] (-z) circle(0.5pt) node [below left] {$-z$};
  \filldraw[black] (phi) circle(0.5pt) node [below=0.4mm] {$\phi$};
  \filldraw[black] (-phi) circle(0.5pt) node [below=0.4mm] {$\pi-\phi$};
  \filldraw[black] (pFalse) circle(1pt) node [below=1.25mm] {$\pi$};
  \filldraw[black] (pTrue) circle(1pt) node [below=1.25mm] {$0$};

  % Comparisons
  \draw [thin, dashed] (z) -- (phi);
  \draw [thin, dashed] (-z*) -- (-phi);

\end{tikzpicture}
\caption{
Comparision of pTruth values of complex numbers on the unit circle. Note the
pTruth of $z=e^{i\phi}$ is the same as that of its complex conjugate with a similar
relationship for $-z$ and $-z^*$. Note also the two domains, $\Theta_1$ marked
by the upper (thick, solid) arc and $\Theta_2$ marked by the lower (thick,
dashed) arc.
}
  \label{fig:compnos}
\end{figure}
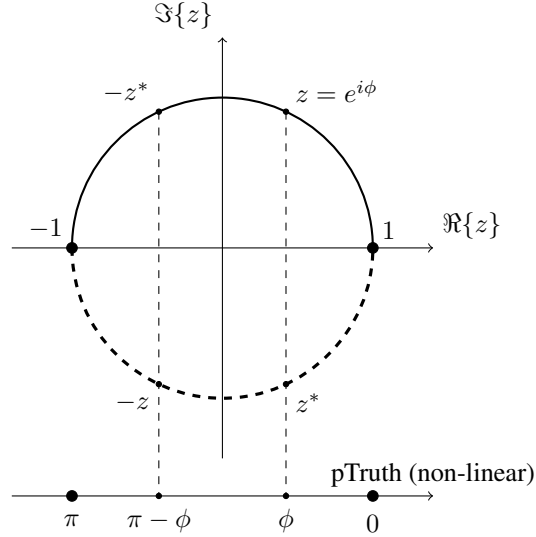

It is obvious that for each $\theta_1 \in \Theta_1$ there exists $\theta_2 \in
\Theta_2$ such that $T_0(\theta_1) = T_0(\theta_2)$ and that $\theta' = 2\pi -
\theta$.  Equivalently the phase-truth of some complex number $z$ is identical
to that of its complex conjugate $z^*$ as illustrated in Figure
\ref{fig:compnos}.

The representation of phase in the NMR system is implemented in a way that
could be combined with the magnitude implementation described above. Again,
starting from equilibrium, a $(-\frac{\pi}{2})_y$ pulse is applied so that \vM is
aligned along the positive $x$-axis. Now the vector is viewed in a rotating
frame offset from the on-resonance frame by a frequency
$\frac{\omega_\mathrm{off}}{2\pi}$ .

In this frame the vector is allowed to precess for a time $\tau_\mathrm{rd}$,
such that for a desired phase $\theta$

\begin{equation}
\omega_\mathrm{off}\tau_\mathrm{d} = \theta.
\end{equation}

\begin{figure}[h!tb]
\centering
\includegraphics{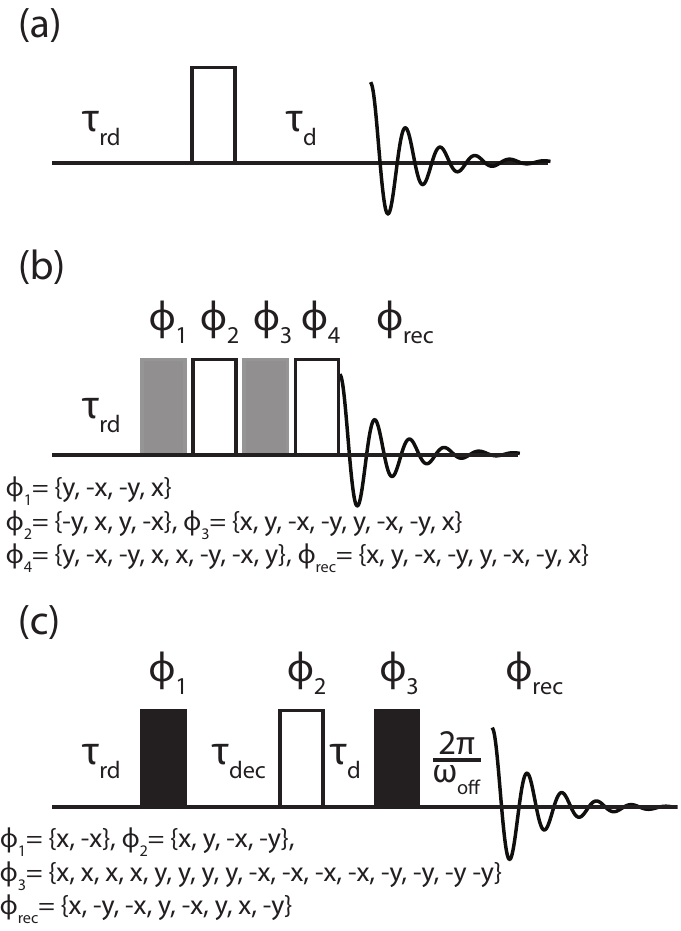}
\caption{Pulse sequences used for complex logic experiments. The experiments in
(a) and (b) are both used to encode phase while that of (c) can encode magnitude
and phase. The presence of multiple pulses in (b) and (c) require repetition
with phase cycling as shown below sequence to compensate for instrumental
imperfections. Grey pulses are of variable duration, unfilled pulses correspond
to $\frac{\pi}{2}$ pulses and black to $\pi$ rotations.}
\label{fig:comppulse}
\end{figure}

The encoded phase is computed from the Fourier transform of the measured time
domain signal. The pulse sequence is shown in Figure \ref{fig:comppulse}(a).  We
also encode the phase in the on-resonance frame, using a pulse sequence
described in Figure \ref{fig:comppulse}(b). The former method was found to be
broadly more reliable.

As with mLogic, it is defined that a pLogic operation (denoted $F_p$)
acting on some complex number is related to the fuzzy logic equivalent. However,
it is not a simple case of writing $F_p(re^{i\theta}) = r\exp(iF_0(\theta))$
since the fuzzy operations do not reflect the new truth-values (as defined
above).

We define an operation in fuzzy logic $F$ to match a function in pLogic
$F_\mathrm{peq}$ under the condition that

\begin{equation}
T_0(F_\mathrm{peq}(\theta_1, \cdots \theta_n)) = F_0(T_0(\theta_1),
\cdots T_0(\theta_n)).
\label{eqn:peqdef}
\end{equation}

An operation from fuzzy logic can then be described in pLogic by

\begin{equation}
F_p(re^{i\theta}) = r \exp(iF_\mathrm{peq}(\theta)).
\label{eqn:pfnct}
\end{equation}

Mirroring the behaviour of mLogic, the phase operations alone do not define
the magnitude of the output in a non-unary operation. For example, pXNOR
\begin{equation}
r_1e^{i\theta_1} \XNOR_p r_2e^{i\theta_2} =
r_\mathrm{res}e^{i(\theta_1 + \theta_2)},
\label{eqn:pXNOR}
\end{equation}
where $r_\mathrm{res} = r_\mathrm{res}(\theta_1, \theta_2)$ is not defined as in
equations \ref{eqn:generalop} and \ref{eqn:mAND}.

\subsection{Combined Complex-Number-Based Logic}

\label{sec:total}

It just so happens that when complex number multiplication is substituted for
$f$ in equation \ref{eqn:generalop}, ($f(z_1, z_2) = z_1 \times z_2$), the
resulting expression gives an mAND and a pXNOR gate

\begin{align}
g(r_1, r_2) &= r_1 \times r_2 = r_1 \wedge r_2 \\
h(\theta_1, \theta_2) &= \theta_1 + \theta_2 \mod 2\pi = \theta_1 \XNOR
\theta_2 \\
z_1 \times  z_2 &= \underbrace{r_1 r_2}_{\text{mAND}}
\exp({i(\underbrace{\theta_1 +\theta_2)}_{\text{pXNOR}}}).
\end{align}

We thus have the operations necessary for a half adder \cite{Tanenbaum1998}.

This is just one example of many possible \emph{Complex Logic Gates}
that combine magnitude and phase logic.

In order to implement such a gate it is necessary to combine the magnitude and
phase implementation as described above. In this process, \vM is manipulated to
encode magnitude exactly as before but an additional precession time is allowed
before the measurement. In this way the magnetisation vector in the $x$-$y$
plane has both magnitude and phase encoded in it. The process is depicted in
Figure \ref{fig:comppulse}(c).

This process is used to implement complex number multiplication: by choosing
$\tau_{dec}$ and $\tau_d$ appropriately it was possible to position \vM in the
x-y plane at any position. Arbitrary scaling of the maximum length of \vM allows
any magnitude to be chosen.

Such experiments are highly accurate, with the magnitude accurate to five parts
in one thousand and the phase to one part in one hundred. This is demonstrated
in Figure \ref{fig:postergraphs}.

%% file: sections/experimental.tex
\subsection{Spectrometer}

All $\mathrm{^1H}$ NMR experiments were performed on a Bruker Avance II 700 NMR
spectrometer (corresponding to a $\mathrm{^1H}$ Larmor frequency
$\frac{\omega_0}{2\pi} = \SI{-700.13}{\mega\hertz}$) equipped with a commercial
triple-resonance ($\mathrm{^1H}$ / $\mathrm{^{13}C}$ / $\mathrm{^{15}N}$) probe
at $T = \SI{289}{\kelvin}$. Samples were contained in standard
$\SI{5}{\milli\meter}$ o.d.  NMR tubes.  Durations of (calibrated) non-selective
$\frac{\pi}{2}$ pulses were of the order of $\SI{7.5}{\micro\second}$.

\subsection{Samples and Experiments}

$\mathrm{^1H}$ NMR experiments on ternary logic were carried out on a sample of
$99.9$ percent deuterated \ce{CDCl3} to which a small amount of \ce{H2O} was
added. This sample provided two well separated $\mathrm{^1H}$ resonances
originating from \ce{H2O} and residual \ce{CHCl3} (see Figure
\ref{fig:selectdelay}).  Relaxation delays of \SI{5}{\second} were found to be
sufficient. Selective excitation experiments were performed using Gaussian
excitation profile pulses of duration $\SI{4.244}{\milli\second}$ corresponding
to a $\frac{\pi}{2}$ rotation.

All $\mathrm{^1H}$ NMR experiments on complex number based logic were carried
out on the $\mathrm{^1H}$ NMR resonance of residual \ce{CHCl3} in a sample of
$99.9$ percent deuterated \ce{CHCl3}. The sample was contained in an NMR tube
fitted with a J. Young valve. The $\mathrm{^1H}$ $T_1$ value for the sample
was determined to be $\SI{7.6+-0.1}{\second}$, using standard inversion
recovery\cite{Levitt2008}. Applying standard phase cycling, precision of
phase and magnitude outputs were found to be one part in one thousand and one
part in ten thousand respectively.

\begin{figure}[h!tb]
\centering
\subfloat[]{
\includegraphics[width=\linewidth]{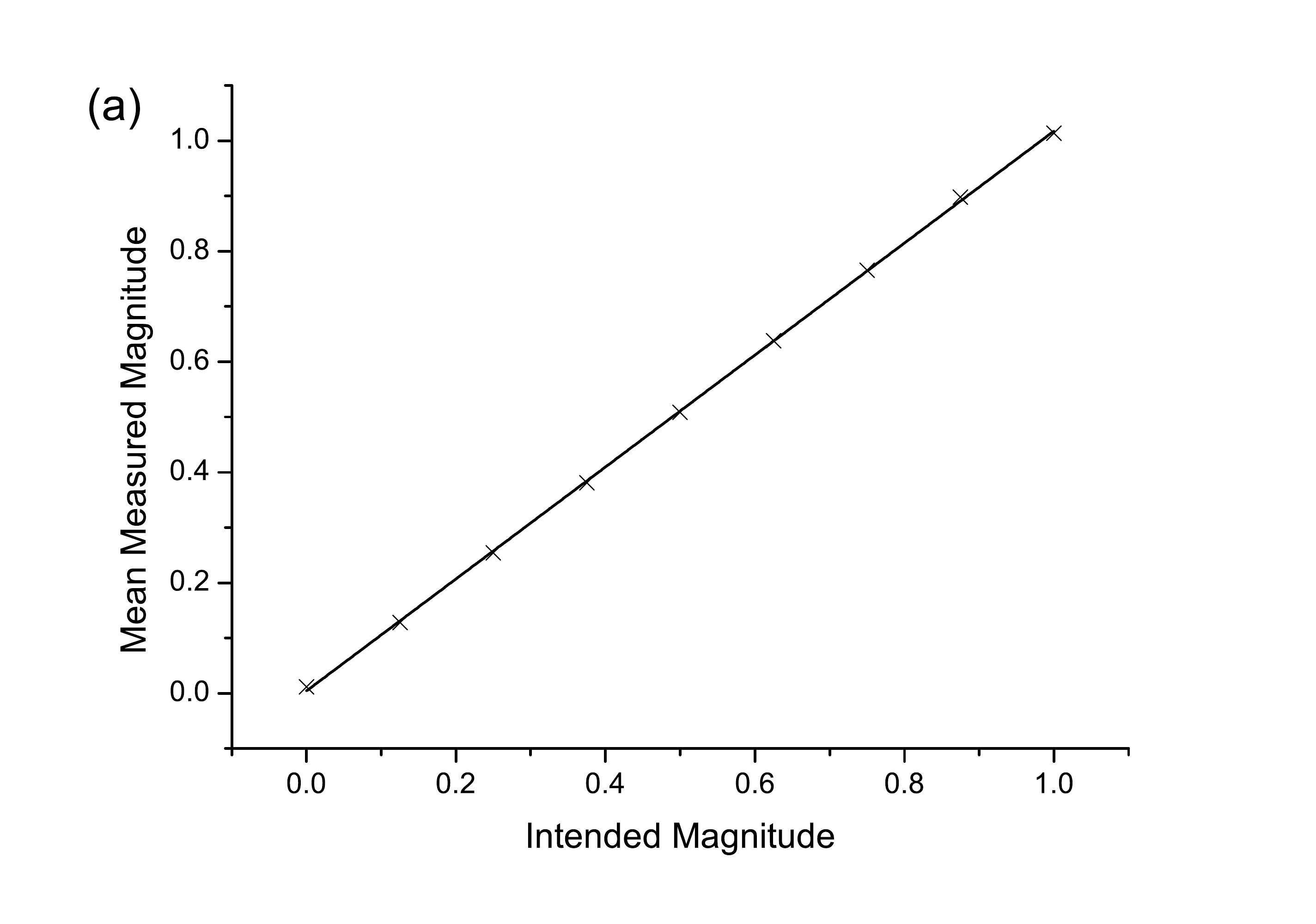}
\label{subfig:postermag}
} \\ \vspace{-2cm}
\subfloat[]{
\includegraphics[width=\linewidth]{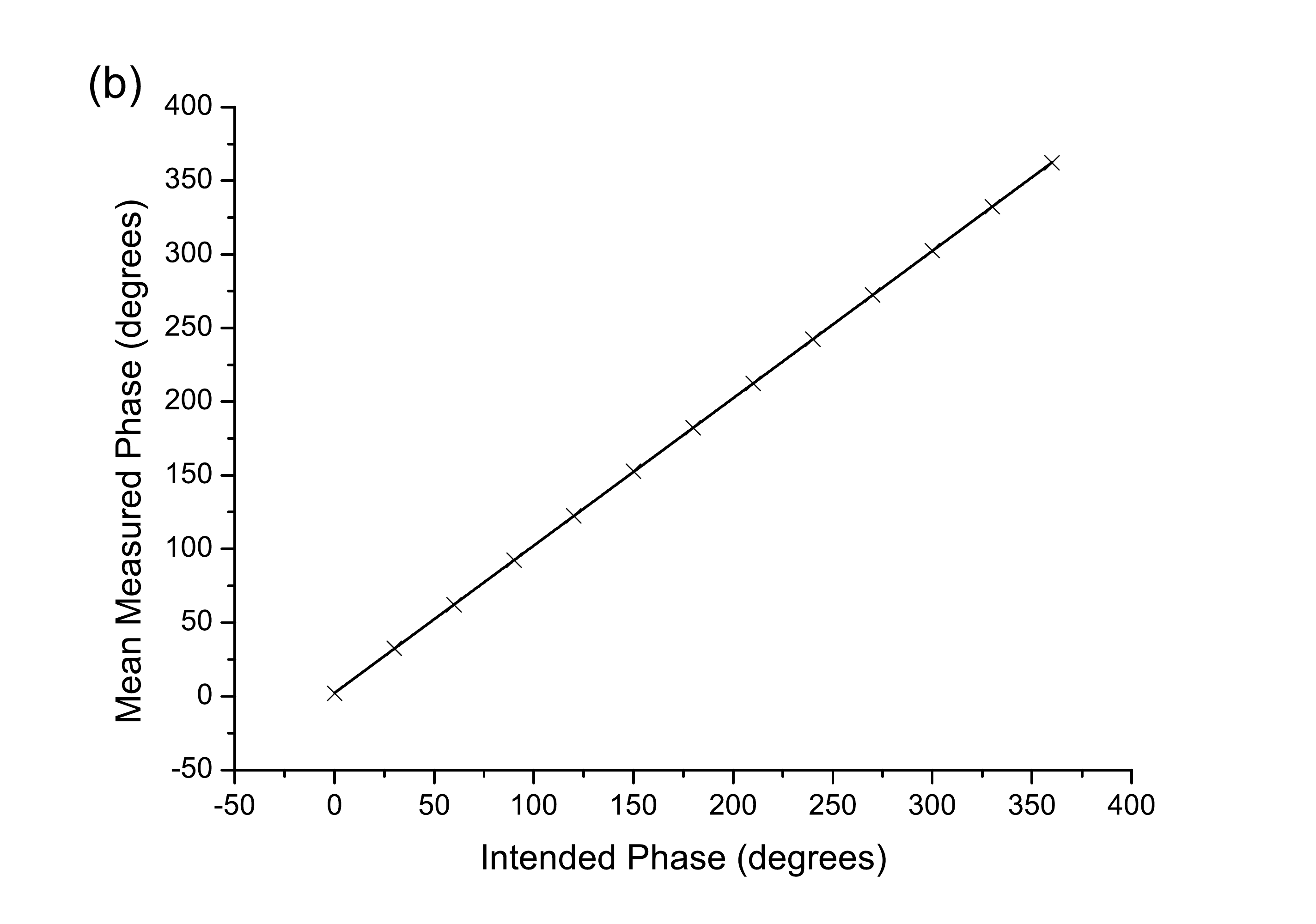}
\label{subfig:posterpha}
}
\caption{Complex number multiplications using the phase sequence in Figure
\ref{fig:comppulse}(c).  \refsubonly{subfig:postermag} and
\refsubonly{subfig:posterpha} show the mean measured result vs the anticipated
for magnitude and phase respectively.}
\label{fig:postergraphs}
\end{figure}

Simulations of the NMR experiments used to implement ternary logic gates were
created using Matlab R2012b (8.0.0.783).

%% file: sections/outlook.tex
In the light of our experimental implementations it should be abundantly clear
that we do not advertise NMR-based computation as a particularly practical
approach. However, nuclear spin dynamics are eminently rich and controllable
and, thus, ideally suited as a design tool and sandpit for all kinds of
exploration and scrutiny when developing new ideas in logic gates and circuitry.
This is particularly true in that a wide variety of implementations can be
tested on the same system and will allow for direct comparisons of concepts of
cost, robustness, or error-propagation behaviour.

As far as NMR systems are concerned we have so far barely scratched the surface
of the richness of its parameter space: we have solely considered nuclear spin
dynamics in small molecules at ambient conditions, in a strong and homogeneous
external magnetic field in the presence of rapid isotropic molecular tumbling
and in the absence of internuclear dipolar coupling interactions. This narrow
window provides particularly straightforward experimental conditions as the
underlying nuclear spin dynamics can be fully described by a vector picture
\cite{Levitt2008}.

If more complicated nuclear spin dynamics ``hardware'' is required, it will be
extremely straightforward to exploit more of the capabilities of NMR systems.
For example, using an additional  magnetic field gradient across the volume of a
single-component sample will encode a range of Larmor frequencies over the
sample volume, similar to the techniques used for spatial encoding in medical
applications of magnetic resonance \cite{Levitt2008}. Such manipulations may
have attractive features regarding parallel operations or in the creation of
dynamic memory.

One could alternatively use chemical samples of a more complicated make-up, with
multiple different sites and isotopes (and corresponding resonance frequencies).
Potentially this could include internuclear dipolar couplings and could be
combined with more sophisticated techniques, such as those found in
two-dimensional NMR experiments \cite{Levitt2008}.

While there may be rather obvious potential benefits from increasingly
complicated nuclear spin dynamics, the drawback is that for many such systems
one will have to employ numerical simulations to predict and engineer the exact
behaviour of the NMR system. In itself this is not a problem but it has some
impact on the role of NMR systems as a sandpit for developmental work for
unconventional computations that could subsequently be implemented on other
physical systems. While numerically exact simulations are perfectly feasible for
even quite complicated nuclear spin systems and NMR experiments
\cite{Levitt2008}, this option does not normally exist to the same degree for
other physical systems.

This is particularly relevant when seeing the (spin based) NMR system as a
development and checking tool for other spin-based methods, say optical
computation. Optical computation is far more likely to eventually become a
practical computational approach and may benefit from developmental work using
NMR systems --- as long as there is no need for numerical simulations.

Finally we mention in passing that one of the features that renders NMR
computation less practical is the timescales of operation: even under favourable
conditions such as we used here in solution-state NMR, relaxation times for
nuclear spin ensembles to return to equilibrium magnetisation (equivalent to a
refreshed system) are of the order of several to many tens of seconds (and can
be much longer in other forms of condensed matter).  However, the natural
slowness of the system, together with the timescales of r.f. pulses of the order
of $\mu \mathrm{s}$, allows for the design of unusual systems in which many
computational steps are reversible to a very good degree of approximation.

%% file: sections/ternaryclasses.tex
\label{ternaryclasses}

Two ternary logic functions which take two inputs and give one output, $f_1(x,y)$ and $f_2(x,y)$, are considered to be in the same Negation-Permutation-Negation (NPN) class if they are equivalent under any of three definitions:

Equivalence A: $f_1(x,y)$ is equivalent to $f_2(x,y)$
if $f_1 = P_i(f_2)$ for any of the permutation functions, $P_i$, of the set
$(0,1,2)$. That is, permuting the output of one function leads to another
function in the same class.

Equivalence B: The two functions are equivalent if
$f_1(x,y) = f_2(P_i(x), P_j(y))$. That is, permuting the input values of a
function leads to a function in the same class.

Equivalence C: The two functions
are equivalent if $f_1(x,y) = f_2(y,x)$. That is, swapping the order of the
inputs gives another function in the same class.

These equivalences are illustrated in Figure \ref{fig:class33}.

%% file: sections/generalisation.tex
\label{app:general}

pLogic as defined in section \ref{complex} is only one possible reinterpretation
of continuous logic and may not be suitable for implementation in other systems.

Define a \emph{Continuous Logic} of order $n$ as a 4-tuple

\begin{equation}
\label{deflog}
L = (\Lambda, I, B, \preceq).
\end{equation}

\begin{itemize}
\item $\Lambda$ is the \emph{Logical Domain}, an interval.

\item $I\subset\Lambda$ is a tuple of the $n$ \emph{absolutes} with
$I =(\absi{0}, \absi{1}, \cdots \absi{n-2}, \absi{n-1}).$ These are
generalisations of the concepts of ``True'' and ``False.''

\item $B$ is the set of \emph{base gates}, the distinct functions that
will act on members of $\Lambda$.

\item Implicitly defined is $G$, the \emph{set of all possible gates} $G$ where
for any $g \in G$, $g$ is some partial composition of any number of members of
$B$.

\item $\preceq$ is a transitive, reflexive, antisymmetric, binary function that
\emph{partially orders} $G$, with the additional requirement that
for all $k, k+1 \in I$ and all $\absi{j} \in I$ we have $\absi{k} \preceq
\absi{k+1}$
% TODO Does it need to be a strict ordering?
\end{itemize}

The properties of any base gate $b \in B$ of order $m$ are as follows
\begin{enumerate}
\item $b:\Lambda^m \rightarrow \Lambda$
\item it is not possible to construct the function $b$ from some partial
composition of other members of $B$.
\end{enumerate}

For example, conventional fuzzy logic is described by

\begin{equation}
\label{fuzzyformal}
L_\mathrm{Fuzzy} = ([0,1], (0,1), B_\mathrm{Fuzzy},  \leq)
\end{equation}

for an appropriate choice of $B_\mathrm{Fuzzy}$ such as XNOR and fanout.

A logic is said to be \emph{gate-complete} under the condition that $G$ is a set
of all possible functions that can act on the domain. Equivalently, that $B$
forms a set of \emph{universal gates}.

Now consider any two logics $L = (\Lambda, I, B, \preceq)$ and $L' = (\Lambda',
I', B', \preceq')$ with corresponding $G$ and $G'$ respectively. We now state
that all $\lambda \in \Lambda$, $\iota \in I$, $g \in G$ and equivalently for
the primed logic $\lambda' \in \Lambda'$, etc.

% TODO The projecting function might actually be a RELATION
We say that $L$ \emph{projects} to $L'$ under a surjective function $T: \Lambda
\rightarrow \Lambda'$ if for all $\absi{k}$ and $\absi{k}'$ we have

\begin{equation}
T(\absi{k}) = \absi{k}' %T discontinuities at i only?
\end{equation}

$T$ is known as the \emph{projecting function} and any $\lambda$ and
$\lambda'$ with $T(\lambda) = \lambda'$ are called \emph{equivalent truth
values}.

For any such logics, a function $g$ can be said to \emph{weakly match} a function
$g'$ under the condition that

\begin{equation}
T(g(\absi{1}, \cdots \absi{m})) = g'(T(\absi{1}), \cdots T(\absi{m}))
\end{equation}

which is denoted

\begin{equation}
g \sim_T g'
\end{equation}

Furthermore, a function $g$ can be said to \emph{strongly match} some $g'$ if

\begin{equation}
T(g(\lambda_{1}, \cdots \lambda_{m})) = g'(T(\lambda_{1}), \cdots T(\lambda_{m}))
\end{equation}

which is denoted

\begin{equation}
g \simeq_T g'
\end{equation}

For any $L$ and $L'$, $L$ \emph{weakly corresponds} to $L'$ under $T$ if

\begin{enumerate}
\item $L$ projects to $L'$ under $T$
\item For each $g$ there exists some $g'$ such that $g \sim_T g'$
\end{enumerate}

in which case we write

\begin{equation}
L \approx_T L'.
\end{equation}

$L$ \emph{strongly corresponds} to $L'$ under $T$ if

\begin{enumerate}
\item $L$ projects to $L'$ under $T$
\item For each $g$ there exists some $g'$ such that $g \simeq_T g'$
\end{enumerate}

in which case we write

\begin{equation}
L \approxeq_T L'.
\end{equation}

It can then be shown that pLogic weakly corresponds to Fuzzy Logic under
projecting function $T_0$ as given in equation \ref{eqn:proj} and the sets of
functions contain only the NOT, XOR and XNOR gates. It might be possible to
define further gates that will match in the two systems but that is not
discussed here.

A final point is on the intervals of the logical domains that are between any
two absolutes, which we shall call \emph{absolute intervals}. In pLogic, there
are two such domains, $\Theta_1 = (0,\pi)$ and $\Theta_2 = (\pi, 2\pi)$ as
depicted in figure \ref{fig:compnos}. This maps exactly to binary continuous
logic, since each truth value corresponds to one in the interval $(0,1)$.

\begin{figure}[h!]
\centering
\begin{tikzpicture}[scale=2.0]
  % Axes
  \draw [->] (-1.4,0) -- (1.4,0) node [above right] {$\Re\{z\}$};
  \draw [->] (0,-1.4) -- (0,1.4) node [above left] {$\Im\{z\}$};

  % Points
  % Circle
  \coordinate (O) at (0,0);
  \coordinate (1) at (1,0);
  \coordinate (B) at (120:1);
  \coordinate (C) at (240:1);
  % Draw nodes
  \filldraw[black] (1) circle(1pt) node [above right] {\absi{0}};
  \filldraw[black] (B) circle(1pt) node [above left] {\absi{1}};
  \filldraw[black] (C) circle(1pt) node [below left] {\absi{2}};

  % Circle
  \draw (1) arc (0:120:1);
  \draw[thick, dashed] (B) arc (120:240:1);
  \draw[thick] (C) arc (240:360:1);

  % Label intervals
  \draw (0.8,0.85) circle(0pt) node {$\Theta_0$};
  \draw (-1.15,0.2) circle(0pt) node {$\Theta_1$};
  \draw (0.8,-0.85) circle(0pt) node {$\Theta_2$};

\end{tikzpicture}
\caption{
A variation on phase logic containing three absolutes and hence three absolute
intervals, $\Theta_0$, $\Theta_1$ and $\Theta_2$ which are marked by solid,
dashed and thick lines respectively.}
\label{fig:general:domains}
\end{figure}
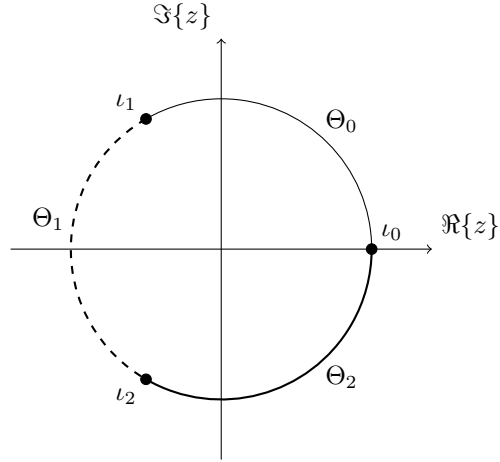

However, if pLogic is modified so that, for example, $I = \{0, \frac{2\pi}{3},
\frac{4\pi}{3}\}$ we will have a continuous ternary logic (as opposed to the
discrete ternary logic described in Section \ref{ternary}. There are now three
absolute intervals: $(0, \frac{2\pi}{3})$, $(\frac{2\pi}{3},\frac{4\pi}{3})$ and
$(\frac{4\pi}{3}, 0)$. Such a system is depicted in Figure
\ref{fig:general:domains}.

One might think that this could trivially map to some corresponding modification
of binary continuous logic with $I = \{0,\frac{1}{2}, 1\}$ but that is not the
case as there are now only two absolute intervals: $(0,\frac{1}{2})$ and
$(\frac{1}{2},1)$. It would therefore be impossible to describe a truth-value
that is somewhere between $1$ and $0$ whereas this can be done in the modified
pLogic.

We suggest that by a suitable generalisation of the work here it  may be
possible to describe any $n^\mathrm{th}$ order logic with absolute domains for
any two absolutes in the Logic.

% TODO
% Domains
% Inductive gates?
% Hint at future things to do

%% file: nmrlogic.bbl
\begin{thebibliography}{10}

\bibitem{Amos2002}
M.~Amos, G.~Paun, G.~Rozenberg, and A.~Salomaa.
\newblock (2002).
\newblock Topics in the theory of {DNA} computing.
\newblock {\em Theoretical Computer Science}, 287(1):3 -- 38.
\newblock Natural Computing.

\bibitem{Whiting2014}
J.~G. Whiting, B.~P. de~Lacy~Costello, and A.~Adamatzky.
\newblock (2014).
\newblock Slime mould logic gates based on frequency changes of electrical
  potential oscillation.
\newblock {\em Biosystems}, 124:21--25.

\bibitem{Gorecki2015}
J.~Gorecki, K.~Gizynski, J.~Guzowski, J.~Gorecka, P.~Garstecki, G.~Gruenert,
  and P.~Dittrich.
\newblock (2015).
\newblock Chemical computing with reaction--diffusion processes.
\newblock {\em Phil. Trans. R. Soc. A}, 373(2046):20140219.

\bibitem{Harding2005}
S.~Harding and J.~F. Miller.
\newblock (2005).
\newblock Evolution in materio: Evolving logic gates in liquid crystal.
\newblock In {\em In Proceedings of the workshop on unconventional computing at
  ECAL 2005 VIIIth European}, page~12.

\bibitem{Swade2002}
D.~Swade.
\newblock (2002).
\newblock {\em The Difference Engine: Charles Babbage and the Quest to Build
  the First Computer}.
\newblock Penguin Books.

\bibitem{Murrell2013}
K.~Murrell, D.~Holroyd, National{\ }Museum{\ }of{\ }Computing, and National{\
  }Museum{\ }of{\ }Computing{\ }Staff.
\newblock (2013).
\newblock {\em The Harwell Dekatron Computer}.
\newblock National Museum of Computing.

\bibitem{Bechmann2012}
M.~Bechmann, A.~Sebald, and S.~Stepney.
\newblock (2012).
\newblock Boolean logic gate design principles in unconventional computers: an
  nmr case study.
\newblock {\em International Journal of Unconventional Computing},
  8(2):139--159.

\bibitem{Jones2011}
J.~A. Jones.
\newblock (2011).
\newblock Quantum-computing with {NMR}.
\newblock {\em Progress in Nuclear Magnetic Resonance Spectroscopy},
  59:91--120.

\bibitem{Glusker2005}
M.~Glusker, D.~Hogan, and P.~Vass.
\newblock (July 2005).
\newblock The ternary calculating machine of {T}homas {F}owler.
\newblock {\em Annals of the History of Computing, IEEE}, 27(3):4--22.

\bibitem{Brusentsov2011}
N.~P. Brusentsov and J.~R. Alvarez.
\newblock (2011).
\newblock Ternary computers: The setun and the setun 70.
\newblock In J.~Impagliazzo and E.~Proydakov, editors, {\em Perspectives on
  Soviet and Russian Computing}, volume 357 of {\em IFIP Advances in
  Information and Communication Technology}, pages 74--80. Springer Berlin
  Heidelberg.

\bibitem{Levitt2008}
M.~H. Levitt.
\newblock (2008).
\newblock {\em Spin Dynamics: Basics of nuclear magnetic resonance}, 2 edition.
\newblock John Wiley \& Sons, Ltd. Chichester.

\bibitem{Correia2001}
V.~P. Correia and A.~Reis.
\newblock (2001).
\newblock Classifying n-input boolean functions.
\newblock In {\em in Proc. IWS}, page~58.

\bibitem{Lloris1981}
A.~Lloris and J.~Velasco.
\newblock (July 1981).
\newblock Classification under permutations of the ternary functions of two
  variables.
\newblock {\em Computers and Digital Techniques, IEE Proceedings E},
  128(4):143--148.

\bibitem{Soma2011}
T.~Soma and T.~Soma.
\newblock (May 2011).
\newblock Classification of ternary logic functions by self-dual equivalence
  classes.
\newblock In {\em Multiple-Valued Logic (ISMVL), 2011 41st IEEE International
  Symposium on}, pages 33--37.

\bibitem{Levin2000}
V.~Levin.
\newblock (2000).
\newblock Continuous logic i. basic concepts.
\newblock {\em KYBERNETES}, 29(9-10):1234--1249.

\bibitem{Tanenbaum1998}
A.~S. Tanenbaum and J.~R. Goodman.
\newblock (1998).
\newblock {\em Structured Computer Organization}, 4th edition.
\newblock Prentice Hall PTR, Upper Saddle River, NJ, USA.

\end{thebibliography}
